\documentclass{article}

\usepackage{PRIMEarxiv}

\usepackage[utf8]{inputenc} 
\usepackage[T1]{fontenc}    
\usepackage{hyperref}       
\usepackage{url}            
\usepackage{booktabs}       
\usepackage{amsfonts}       
\usepackage{nicefrac}       
\usepackage{microtype}      
\usepackage{lipsum}
\usepackage{fancyhdr}       
\usepackage{graphicx}       
\usepackage{algorithmic}
\usepackage{algorithm}
\usepackage{mathtools}
\usepackage{array}
\usepackage[caption=false,font=normalsize,labelfont=sf,textfont=sf]{subfig}
\usepackage{textcomp}
\usepackage{stfloats}
\usepackage{verbatim}
\usepackage{color}
\usepackage{threeparttable}
\usepackage{cite}
\graphicspath{{media/}}     
\newcommand{\argmin}{\operatornamewithlimits{argmin}}

\title{Diff-INR: Generative Regularization for Electrical Impedance Tomography}

\author{
  Bowen Tong, Junwu Wang, Dong Liu\\
  University of Science and Technology of China, Hefei, China \\
  \texttt{\{bowentong, wangwuj\}@mail.ustc.edu.cn} \\
  \texttt{dong.liu@outlook.com} \\
}

\begin{document}
\maketitle

\begin{abstract}
  Electrical Impedance Tomography (EIT) is a non-invasive imaging technique that reconstructs conductivity distributions within a body from boundary measurements. However, EIT reconstruction is hindered by its ill-posed nonlinear inverse problem, which complicates accurate results. To tackle this, we propose Diff-INR, a novel method that combines generative regularization with Implicit Neural Representations (INR) through a diffusion model.
  Diff-INR introduces geometric priors to guide the reconstruction, effectively addressing the shortcomings of traditional regularization methods. By integrating a pre-trained diffusion regularizer with INR, our approach achieves state-of-the-art reconstruction accuracy in both  simulation and experimental data. The method demonstrates robust performance across various mesh densities and hyperparameter settings, highlighting its flexibility and efficiency.
  This advancement represents a significant improvement in managing the ill-posed nature of EIT. Furthermore, the method's principles are applicable to other imaging modalities facing similar challenges with ill-posed inverse problems. 
\end{abstract}


\section{Introduction}
Electrical Impedance Tomography (EIT) is a non-invasive, radiation-free soft-field imaging \cite{soleimani2016super} technique with promising applications in medical diagnosis \cite{huaiwu2024_3deit}, \cite{mauri2020covid19}, and industrial process monitoring \cite{wei2016eit_application}. EIT uses electrodes placed on the body's surface to apply small currents and measure the resulting voltages, which are then used to create images of the body's internal structures.
EIT is a valuable tool for monitoring a variety of physiological functions, such as lung ventilation and perfusion \cite{leonhardt2012eit_vq}, as it provides real-time, non-invasive imaging of the electrical conductivity distribution within the body.

Mathematically, EIT reconstruction is a severely ill-posed nonlinear inverse problem, presenting a significant challenge for achieving accurate results. To tackle this issue, data-driven supervised neural networks (NNs) have been extensively employed. These models learn the complex mapping between measurement data and the target image, or they learn post-processing techniques to enhance the quality of the reconstructed image \cite{chen2021mlae_eit}, \cite{hamilton2018deep_dbar}, \cite{seo2019learning_based_eit}. Despite this, the ill-posed nature of EIT makes this mapping highly intricate. Additionally, the limited availability of training data presents a significant obstacle. 
Meanwhile, measurement error and modeling uncertainties significantly increase the data requirements and challenge the NN's capability to handle different situations within a supervised learning framework.

To improve the generalization ability of NNs, self-supervised or unsupervised image reconstruction algorithms have gained increasing attention and achieved promising results in the field of medical imaging.
One prominent example of these approaches is Deep Image Prior (DIP) \cite{DIP}, which has shown excellent performance in image reconstruction tasks such as CT \cite{baguer2020dip_ct}, PET \cite{gong2019dip_pet}, and MRI \cite{yoo2021dip_mri}. 
DIP has also shown promise in EIT reconstruction, as demonstrated by the recently reported DeepEIT framework  \cite{deepeit} and \cite{NASDIP_Xiahaoyuan2024}. The core concept involves using a convolutional NN to represent the unknown conductivity distribution, transforming the image reconstruction problem into an optimization problem for the NN parameters. Despite its potential, DeepEIT continues to encounter challenges, including the insufficient recovery of sharp details (high-frequency information) and a relatively slow convergence rate.
To improve frequency preservation and accelerate the reconstruction, implicit neural representations (INRs) with positional encoding have been successfully applied into EIT \cite{junwu2023inr_eit}. This approach achieves better detail preservation and faster convergence compared to the previously established DeepEIT. It highlights the considerable potential of INR as an effective method for EIT reconstruction. 
Notably, these self-supervised reconstruction problems undergo iterative optimization, where the NNs' iterative process inherently performs a type of {\it implicit} regularization, consistent with the findings in recent work \cite{martin2021implicit}.

Beyond {\it implicit} regularization, NNs can also be employed as an {\it explicit} form of regularization. For example, a trained NN can serve as an  {\it explicit} regularizer, as exemplified by the Regularization by Denoising (RED) framework \cite{RED}. RED is a versatile approach that utilizes denoising NNs to regularize inverse problems. 
RED-based image reconstruction methods typically rely on a paired dataset of noisy data and corresponding ground truth, making them well-suited for specific medical imaging tasks, such as sparse-view CT reconstruction \cite{red_svct}, utilizing either experimental or simulated data.
Additionally, generative models, recognized for their ability to produce rich and high-quality images, serve as flexible priors for inverse problems.
Generative Adversarial Networks (GANs), for instance, have been effectively utilized as generative priors in tasks like image restoration \cite{DGP} and MRI reconstruction \cite{GAN_MRI}, \cite{GAN_MRI_2}, \cite{GAN_MRI_3}, showing promising results.
It should be emphasized that these methods rely heavily on high-quality training datasets. However, it is well known that obtaining paired datasets in EIT is challenging. The difficulties include acquiring accurate ground truth data, dealing with variability in tissue properties, and addressing sensitivity to noise and electrode placement.

In recent years, diffusion models \cite{DDPM}, \cite{rombach2022latent_diffusion}, \cite{song2021scorebased} have emerged as a powerful class of generative models, gaining widespread adoption due to their capability to generate high-quality, diverse images and their relative ease of training compared to GANs. Notably, diffusion models have shown exceptional performance in solving general image inverse problems \cite{DPS}, \cite{DDRM}, \cite{PGDM}, as well as in medical image reconstruction \cite{diff_fastMRI_2023tmi}.
In the field of EIT, researchers have also applied diffusion models for EIT \cite{diff_post_eit}.
However, their use has been quite limited, as diffusion models have primarily been employed for post-processing traditional EIT reconstructions based on the Gaussian-Newton (GN) method \cite{diff_post_eit}.

In this work, inspired by Score Distillation Sampling (SDS) \cite{dreamfusion}, \cite{reddiff}, we propose an {\it explicit} plug-and-play (PnP) generative regularization scheme named Diff-INR for EIT reconstruction. To effectively suppress artifacts that are prevalent in classical reconstructions, we leverage INR to introduce implicit regularization. To the best of our knowledge, this is the first application of diffusion model-based regularization to soft-field tomography \cite{soleimani2016super} reconstruction.
Importantly, Diff-INR is not confined to EIT and the extensions to other imaging modalities are mostly straightforward. The PnP generative regularization can also be integrated into various iterative reconstruction methods, such as those DIP-based on approaches.
The main contributions of this work are summarized as follows.

\begin{itemize}
\item {\bf Generative regularization}. Diff-INR addresses the challenging EIT reconstruction by incorporating generative priors learned from 2D images, thus improving reconstruction accuracy without requiring paired dataset.
\item {\bf State-of-the-Art Performance}. Diff-INR, integrating INR and diffusion models, achieves superior performance in EIT reconstruction in terms of perceptual and quantitative metrics.
\item {\bf Improved Robustness}. An adaptive bandwidth control mechanism is introduced to enhance the robustness and repeatability of INR-based EIT reconstructions, making the approach adaptable to different finite element discretization levels.
\end{itemize}

The structure of this paper is organized as follows: Section \ref{sec_bg} provides a brief overview of EIT problems and the background of the proposed method. Section \ref{sec_methods} offers a detailed description of the proposed Diff-INR. Section \ref{sec_imp.details} delves into the implementation specifics of Diff-INR. Section \ref{sec_eval} evaluates the proposed method through experimental data and robustness studies. Finally, Section \ref{sec_conclusion} presents the concluding remarks.

\section{Background}
\label{sec_bg}
In this section, we introduce the forward and inverse problem of EIT. Additionally, we provide an overview of the INR method and the score-based diffusion model used in the Diff-INR framework.

\subsection{EIT Forward Problem and Inverse Problem}
In EIT, current is injected into the measurement domain $\Omega$ through pairs of electrodes following a predefined stimulation and measurement pattern. 
The resulting potential differences between the electrodes are recorded, and these voltage measurements are used to infer the conductivity distribution within the domain. The conductivity distribution is typically discretized using the Finite Element (FE) method on a triangular mesh.
According to the Complete Electrode Model (CEM) \cite{cem_model}, the EIT forward problem is represented by solving the following system of equations for coordinates $C = (x, y)$ within the domain:
\begin{equation}
\label{eq.cem}
\begin{aligned}
\nabla\cdot(\sigma(C)\nabla u(C))& =0,\quad C\in\Omega\\
u(C)+z_q\sigma(C)\frac{\partial u(C)}{\partial\boldsymbol{\nu}}& =U_q,\quad C\in e_q,q=1,\ldots,N_e\\
\int_{e_q}\sigma(C)\frac{\partial u(C)}{\partial\boldsymbol{\nu}}\mathrm{d}S& =I_q,\quad q=1,\ldots,N_e\\
\sigma(C)\frac{\partial u(C)}{\partial\boldsymbol{\nu}}& =0,\quad C\in\partial\Omega\backslash\bigcup_{q=1}^{N_e}e_q\\
\sum_{q=1}^{N_e}I_q = &0, \quad \sum_{q=1}^{N_e}U_q=0.
\end{aligned}
\end{equation}
Here, $\sigma$ represents the conductivity distribution, $u$ denotes the electrical potential, $\boldsymbol{\nu}$ is the normal vector to the boundary, and $I_q$ and $U_q$ refer to the current and voltage at the $q$-th electrode, respectively. 
By numerically approximating the CEM using FE method, the observation model can be expressed as:
\begin{equation}
\label{eq.eitforward}
V = U(\sigma) + \epsilon,
\end{equation}
where $U$ is the forward operator that maps the conductivity distribution $\sigma$ to the measured voltage $V$, and $\epsilon$ represents additive noise.

The EIT inverse problem involves recovering the conductivity distribution $\sigma$ from the voltage measurements $V$ using the forward model. This is typically solved by iteratively optimizing the following minimization problem:
\begin{equation}
\label{eq.inverse} 
\argmin_\sigma \left\{\|V - U(\sigma)\|^2 + \alpha R(\sigma)\right\}. 
\end{equation}
The first term represents the data-fidelity loss, quantifying the difference between the measured data $V$ and the model prediction $U(\sigma)$. The second term introduces a regularization $R(\sigma)$, where $\alpha$ is a hyperparameter controlling the balance between the data fidelity and regularization, ensuring stability during the optimization process.

\subsection{Implicit Neural Representation with Positional Encoding}
INR has gained significant attention in computer vision and image reconstruction tasks due to its flexible and powerful approach to representing unknowns, such as images or conductivity distributions, using a NN model $F$ \cite{rff}, \cite{luigi2023inr_3d}, \cite{inr_sparse_ct_2023tci}. In this method, the NN receives spatial coordinates $\mathbf{C} =(x, y)$ as input and outputs the corresponding function value $F_\theta(x, y)$, where $\theta$ denotes the network's weights.
To improve the network's ability to capture high-frequency details, positional encoding is typically applied to the input coordinates. One common approach for this is Random Fourier Features (RFF) \cite{rff}, which maps the coordinates $\mathbf{C}$ to a high-dimensional space using the following transformation:
\begin{equation}
\label{eq.rff}
\mathbf{C'} = \mathcal{M}_s(\mathbf{C}) = \begin{bmatrix}
\sin(2\pi \, \mathbf{C} \cdot \mathbf{B}(b)) \\
\cos(2\pi \, \mathbf{C} \cdot \mathbf{B}(b))
\end{bmatrix},
\end{equation}
where $\mathcal{M}_s(\mathbf{C})$ represents the RFF mapping, $\mathbf{B}(b) \in \mathbb{R}^{2 \times n}$ is sampled from a normal distribution $\mathcal{N}(0, b^2)$, and $b$ is the bandwidth parameter.

Positional encoding plays a critical role in balancing the neural network's ability to memorize the data and its capacity to generalize effectively \cite{rethink_pe}. In RFF, the bandwidth $b$ controls this balance. Larger $b$ values enable more accurate data representation, but they can reduce the model's generalization ability. Consequently, selecting an optimal $b$ for a specific application often requires multiple iterations to strike the right balance between fidelity and generalization.

\subsection{Score-based Diffusion Models}
Diffusion models have emerged as powerful generative models, capable of producing high-quality images through a two-step process: a forward diffusion and a reverse denoising process. A notable and widely adopted variant is the score-based diffusion model, which is formulated using the Variance Preserving Stochastic Differential Equation (VP-SDE) framework \cite{song2021scorebased}. The forward diffusion process within this VP-SDE framework is governed by the following SDE:
\begin{equation}
\label{eq.f_sde}
\mathrm{d}\mathbf{x} = -\frac{1}{2} \beta(t) \mathbf{x} \mathrm{~d}t + \sqrt{\beta(t)}\mathrm{d}\mathbf{w},
\end{equation}
where $t \in [0, T]$ is the time step index, $\beta(t) = \beta_{\min} + (\beta_{\max} - \beta_{\min}) \frac{t}{T}$ is the noise schedule, $\mathrm{d}t$ is an infinitesimal time increment, $\mathbf{x}$ denotes the image $\mathbf{x}(t)$, and $\mathbf{w}$ represents a standard Wiener process. The image at time step $t$, denoted as $x_t$, is given by $x_t = \sqrt{1 - s_t^2} x_0 + s_t \epsilon$, where $s_t = 1 - e^{-\int_0^t \beta(v) dv}$, and $\epsilon$ is drawn from a standard normal distribution.

The reverse denoising process, which generates samples from the learned distribution, can be defined as:
\begin{equation}
\label{eq.r_sde}
\mathrm{d}\mathbf{x} = -\frac{1}{2} \beta(t) \mathbf{x} \mathrm{~d}t - \beta(t) \nabla_\mathbf{x} \log p_t(\mathbf{x}) \mathrm{~d}t + \sqrt{\beta(t)}\mathrm{d}\bar{\mathbf{w}},
\end{equation}
where $\bar{\mathbf{w}}$ is the reverse Wiener process, and $\nabla_\mathbf{x} \log p_t(\mathbf{x})$ is the score function of the data distribution $S_\phi(\mathbf{x}, t)$. Diffusion models are trained to estimate this score function, allowing the reverse process to sample new data points from the distribution \cite{song2021scorebased}.
As the data distribution $p_t(\mathbf{x})$ is perturbed with Gaussian noise, i.e., $p_t(\mathbf{x}) \sim \mathcal{N}(\mu, s_t^2)$, diffusion models based on VP-SDE can also be trained to predict the reverse noise $\epsilon_\phi(\mathbf{x}, t)$ as in \cite{improved_ddpm}, where:
\begin{equation}
\label{eq.s=epsilon}
S_\phi(\mathbf{x}, t) = -\frac{\mathbf{x} - \mu}{s_t^2} = -\frac{\epsilon_\phi(\mathbf{x}, t)}{s_t}.
\end{equation}

Recent researches have demonstrated that pre-trained diffusion models, $\epsilon_\phi(\mathbf{x}, t)$, provide powerful priors for solving inverse problems \cite{DDRM, DPS, DGP}. These models can also be integrated into PnP frameworks, with the Score Distillation Sampling (SDS) loss \cite{dreamfusion} serving as a regularization term for the inverse problem. This integration avoids the need to compute the Jacobian of the diffusion model $\epsilon_\phi(\mathbf{x}, t)$, thus improving the stability and efficiency of the optimization process \cite{reddiff}.

\begin{figure*}[!t]
\centering
\includegraphics[trim={0 0 0 0},clip,width=6in]{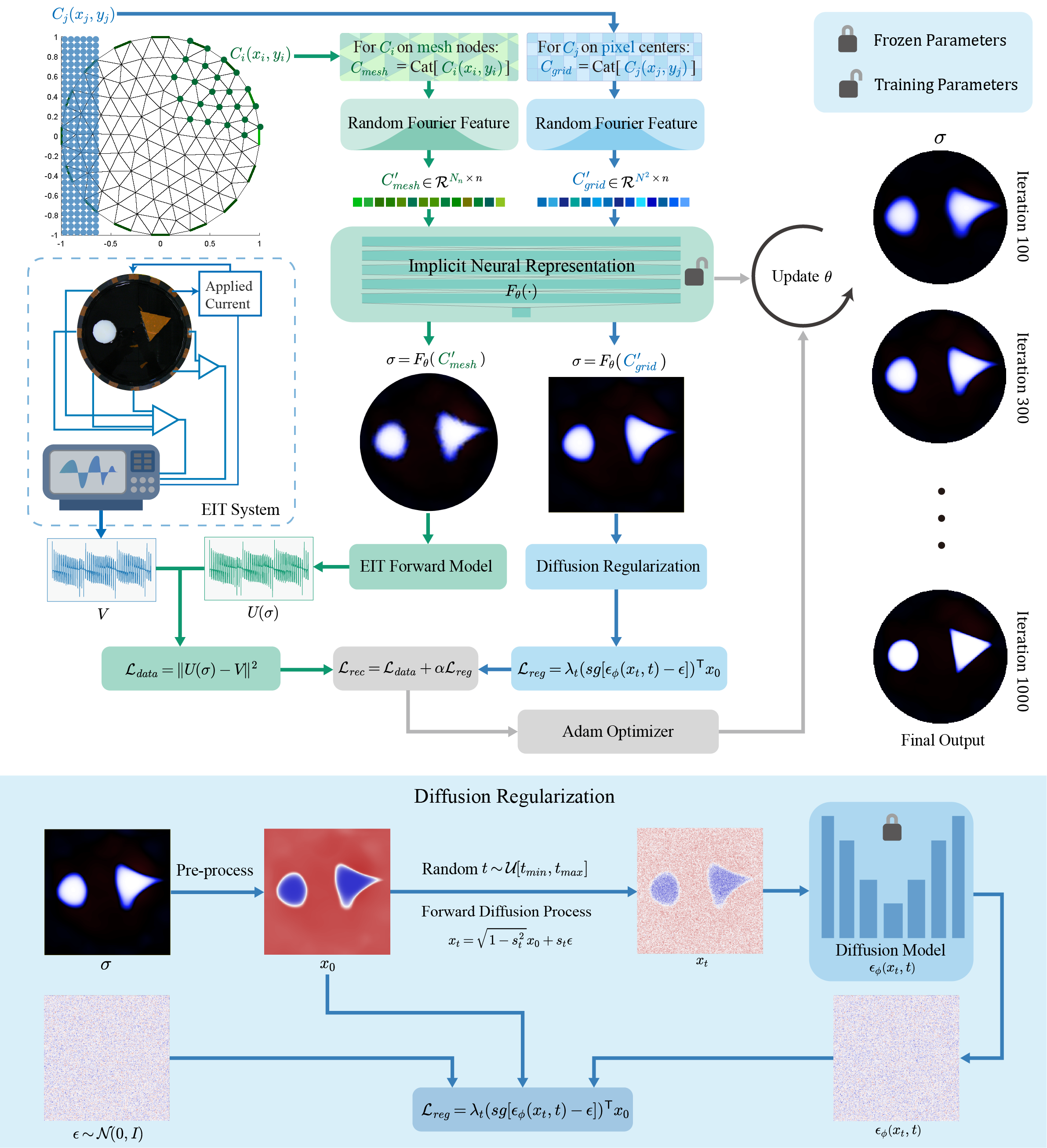}
\caption{Architecture of the proposed method with diffusion regularization. The coordinates of FE nodes and pixel centers are initially encoded using RFF and then fed into the INR to estimate conductivity in both mesh and pixel domains. Data-fidelity loss $\mathcal{L}_\text{data}$ is computed using the EIT forward model, while the regularization loss $\mathcal{L}_\text{reg}$ is calculated based on generative regularization employing a pre-trained diffusion model. The INR is then iteratively optimized with the reconstruction loss $\mathcal{L}_\text{rec}$ using the Adam optimizer until convergence.}
\label{fig.structure}
\end{figure*}

\section{Diff-INR: Implicit Neural Representation with Diffusion Generative Model Regularization}
\label{sec_methods}
This section introduces the proposed Diff-INR method, which integrates generative regularization with the INR framework for EIT reconstruction. We describe the derivation of the generative regularization term, the integration into INR-based reconstruction, and present its corresponding pseudocode. 
The method's architecture is illustrated in Fig.~\ref{fig.structure}.

\subsection{Generative Regularization via Diffusion Model}
\label{sec.diffreg}
In the context of the EIT reconstruction, the objective is to estimate the conductivity distribution $\sigma$ from the measured voltage data $V$. From a variational perspective \cite{reddiff}, the goal is to minimize the Kullback-Leibler (KL) divergence between the variational distribution $q(\mathbf{\sigma}|V)$ and the true posterior $p(\mathbf{\sigma}|V)$, which can be expressed as:
\begin{equation}
\label{eq.variational}
\min_{q(\mathbf{\sigma}|V)}D_{\mathrm{KL}}[q(\mathbf{\sigma}|V)\|p(\mathbf{\sigma}|V)].
\end{equation}
This leads to the following objective:
\begin{equation}
\label{eq.obj}
\min_{q(\mathbf{\sigma}|V)}-\mathbb{E}_{\sigma \sim q(\mathbf{\sigma}|V)}\log{p(V|\mathbf{\sigma})} + D_{\mathrm{KL}}[q(\mathbf{\sigma}|V)\|p(\mathbf{\sigma})] + \log{p(\mathbf{\sigma})}.
\end{equation}
Here, the first term corresponds to the data-fidelity loss:
\begin{equation} \label{eq.l_data} \mathcal{L}_{\text{data}} = || V - U(\sigma) ||^2, \end{equation}
and the second term serves as the regularization term:
\begin{equation} \mathcal{L}_{\text{reg}} \propto D_{\mathrm{KL}}[q(\mathbf{\sigma}|V) | p(\mathbf{\sigma})]. \end{equation}

To compute the $\mathcal{L}_\text{reg}$, 
we propose a generative regularization framework using a diffusion model. A dataset of simple 2D geometric shapes is employed to train a diffusion model that learns the distribution $p(\mathbf{x})$. This distribution is then used to approximate the true conductivity distribution $p(\mathbf{\sigma})$, and the conditioned diffusion posterior $q(\mathbf{x}|V)$ replaces $q(\mathbf{\sigma}|V)$. Following the score-matching method from \cite{song2021scorebased}, the regularization term becomes:
\begin{equation}
\mathcal{L}_\text{reg}\propto \int_0^T\mathbb{E}_{q(x_t|V)}[\lambda(t)\left\|\nabla_{x_t}\log q(x_t|V)-S_{\phi}(x_t,t)\right\|_2^2] \mathrm{d}t,
\end{equation}
where $\lambda(t)$ is a weighting term, and $S_{\phi}(x_t,t)$ is the score function of the pre-trained diffusion model used to approximate $\nabla_{\mathbf{x}}\log p_t(x_t)$.
Using Equation (\ref{eq.s=epsilon}) and the reparameterization trick, this can be simplified to:
\begin{equation}
\label{eq.L_reg_0}
\mathcal{L}_\text{reg} \propto \mathbb{E}_{t \sim \mathcal{U}[0,T],\epsilon \sim \mathcal{N}(0,I)}(\lambda(t)[\epsilon_\phi(x_t, t) - \epsilon]).
\end{equation}

To compute this expectation efficiently, we adopt stochastic optimization by randomly sampling time steps $t$ within a specified range during the optimization. Furthermore, to avoid the computational complexity of calculating the Jacobian of the diffusion model, we apply the SDS approximation \cite{dreamfusion}, \cite{reddiff}, using a stop-gradient (sg) operator,  and the final regularization loss $\mathcal{L}_\text{reg}$ expression becomes:
\begin{equation}
\mathcal{L}_{\text{reg}} = \lambda_t \left(sg[\epsilon_\phi(\mathbf{x_t}, t) - \epsilon]\right)^\mathsf{T} \mathbf{x_0}= \lambda_t \left(sg[\epsilon_\phi(\mathbf{x_t}, t) - \epsilon]\right)^\mathsf{T} F_\theta(\mathbf{C}_{\text{grid}}^\prime),
\end{equation}
where $x_0 = \sigma_\text{grid} = F_\theta(\mathbf{C}_{\text{grid}}^\prime)$. 
Here, $\mathbf{C}_{\text{grid}}^\prime$ represents the positional encoding of every pixel in the grid, aligning the size with that of the pre-trained diffusion model.

With the generative regularization framework established and the regularization loss $\mathcal{L}_\text{reg}$ computed using the diffusion model, we now integrate this into the INR-based reconstruction process. In the following subsection, we outline how the generative regularization is applied within the context of INR-based reconstruction. 

\subsection{Generative Regularization Enabled INR-Based Reconstruction}
To incorporate generative regularization into INR-based reconstruction, we first normalize the reconstruction domain to the domain $x \in [-1, 1]$ and $y \in [-1, 1]$. The pixel coordinates of the image are used for computing the regularization loss $\mathcal{L}_\text{reg}$. In contrast, the coordinates of the FE nodes are used for calculating the data loss $\mathcal{L}_\text{data}$.

Next, we represent the conductivity using an INR function:
\begin{equation}
\sigma_{\text{mesh}} = F_\theta(\mathbf{C}_{\text{mesh}}^\prime) = F_\theta[\mathcal{M}_s(\mathbf{C}_{\text{mesh}})],
\end{equation}
where $\mathbf{C}_{\text{mesh}}$ represents the coordinates of the FE nodes
and $\mathbf{C}_{\text{mesh}}^\prime$ denotes their positional encoding.
Then, the data loss $\mathcal{L}_\text{data}$ can be calculated using:
\begin{equation}
\label{eq.l_data}
\mathcal{L}_\text{data} = \|V - U(F_\theta(\mathbf{C}_{\text{mesh}}^\prime))\|^2.
\end{equation}

Finally, we obtain the loss function of the proposed approach:
\begin{equation}
\label{eq.loss}
\mathcal{L}_\text{rec} = \|V - U(F_\theta(\mathbf{C}_\text{mesh}^\prime))\|^2  + \alpha \lambda_t \left(sg[\epsilon_\phi(x_t, t) - \epsilon]\right)^\mathsf{T} F_\theta(\mathbf{C}_\text{grid}^\prime),
\end{equation}
where $\alpha$ is a regularization weight.

To illustrate the optimization process in detail, we now present the pseudocode, as shown in Algorithm \ref{Alg.Diff-INR}, for the iterative algorithm used to minimize the loss $\mathcal{L}_\text{rec}$ and achieve the final conductivity distribution.

\begin{algorithm}[htb!]
\caption{Diff-INR algorithm for EIT reconstruction}
\begin{algorithmic}[1]\label{Alg.Diff-INR}
\STATE $\textbf{Input:}$ neural network $F_\theta$, measurement voltage $V$, EIT forward model $U$, coordinates of mesh nodes $\mathbf{C}_{mesh}$, coordinates of pixel centers $\mathbf{C}_{grid}$, positional encoding mapping $\mathcal{M}_s$, regularization weight $\alpha$, maximum iterations, and reverse diffusion time step bound $t_{min}$ and $t_{max}$
\STATE $\textbf{Initialize:}$ $\mathbf{C'}_{mesh}=\mathcal{M}_s(\mathbf{C}_{mesh})$, $\mathbf{C'}_{grid}=\mathcal{M}_s(\mathbf{C}_{grid})$
\REPEAT
\STATE $\sigma_{mesh} = F_\theta (\mathbf{C'}_{mesh})$
\STATE $\sigma_{grid} = F_\theta (\mathbf{C'}_{grid})$
\STATE $U = U(\sigma_{mesh})$
\STATE $\mathcal{L}_{data}=\|U-V\|^2$
\STATE Randomly sample $t\sim \mathcal{U}[t_{min},t_{max}]$
\STATE Randomly sample $\mathbf{\epsilon} \sim \mathcal{N}(0, I)$
\STATE $\mathbf{x_0}\gets$ Rescale $\sigma_{grid}$ to $[-1, 1]$
\STATE Compute $\mathbf{x_t}=\sqrt{\alpha_t} \mathbf{x_0} + \sqrt{1-\alpha_t}\epsilon$
\STATE Compute regularization loss $\mathcal{L}_{reg}=\lambda_t(sg[\epsilon_\phi(\mathbf{x_t}, t)-\epsilon])^\mathsf{T} \sigma_{grid}$ $\ \triangleleft$ sg: stop-gradient
\STATE Compute reconstruction loss $\mathcal{L}_{rec} = \mathcal{L}_{data}+\alpha \mathcal{L}_{reg}$
\STATE Compute $\nabla_{\sigma}\mathcal{L}_{data}=2\nabla_{\sigma}U^\mathsf{T} (U-V)$ $\triangleleft$ Jacobian of EIT forward model
\STATE Compute gradients $\nabla_{\theta}\sigma_{mesh}$ and $ \nabla_{\theta}\sigma_{grid}$ using Autograd
\STATE Compute $\nabla_{\theta}\mathcal{L}_{data}={\nabla_{\sigma}\mathcal{L}_{data}}^\mathsf{T} \nabla_{\theta}\sigma_{mesh}$
\STATE Compute $\nabla_{\theta}\mathcal{L}_{reg}= \lambda_t(sg[\epsilon_\phi(\mathbf{x_t}, t)-\epsilon])^\mathsf{T} \nabla_{\theta}\sigma_{grid}$
\STATE Update $\nabla_{\theta}\mathcal{L}_{rec}= \nabla_{\theta}\mathcal{L}_{data} + \alpha \nabla_{\theta}\mathcal{L}_{reg}$
\STATE Update $ \theta $ with Adam optimizer
\UNTIL max-iterations
\end{algorithmic}
\end{algorithm}

\section{Implementation Details}
\label{sec_imp.details}
This section details the implementation specifics of the proposed EIT reconstruction method. It covers the test cases and FE meshes, the training process for the diffusion model, the bandwidth control mechanism for RFF, and execution details. Additionally, we briefly discuss the metrics employed to evaluate the method's performance.

\subsection{Test Cases and Finite Element Mesh Details}
To evaluate the proposed method, we conducted multiple reconstruction tasks using both simulation and experimental data. For the simulation data, we employed a setup that mirrors the experimental configuration. A cylindrical tank with a diameter of 28 cm was simulated, with 16 electrodes equidistantly positioned around the tank. A current of 1 mA at a frequency of 10 kHz was injected into the tank, and the adjacent measurement pattern was employed to collect boundary voltage data. The number of nodes ($N_n$) and elements ($N_e$) for each mesh are listed in Figure \ref{fig.mesh_level}. In the simulated human thorax phantom, the background conductivity was set to 2 mS/cm, while the lung-shaped ellipses were set to 0.5 mS/cm, and the heart-shaped circle was set to 3 mS/cm. To avoid the inverse crime, a finer mesh with $N_n = 5833$ and $N_e = 11424$ was used for the forward problem, which was distinct from the mesh used for reconstruction.

For the experimental data, a tank filled with saline solution was used, containing non-conductive plastic objects of different shapes. Four objects were used in the experiment: a large triangle (31.29 cm²), a small triangle (22.98 cm²), a circle (30.19 cm²), and a rectangle (28.7 cm²). The experiment data was collected using the KIT-4 measurement system \cite{kourunen_kit4}. A total of four experimental cases were selected for comparative analysis with other reconstruction methods.

\subsection{Diffusion Model Training}
In order to introduce shape-guided regularization in the EIT reconstruction process, we designed a dataset consisting of overlapping 2D geometric shapes. These shapes included regular geometries such as circles, triangles, rectangles, and polygons, as well as irregular shapes defined by closed B\'ezier curves. The conductivity values of each shape and the background were randomly assigned within a certain range, allowing the shapes to overlap and form more complex compositions. This ensures the method's generalization capability.

\begin{figure}
    \centering
    \includegraphics[trim={80 60 0 60},clip,width=3.4in]{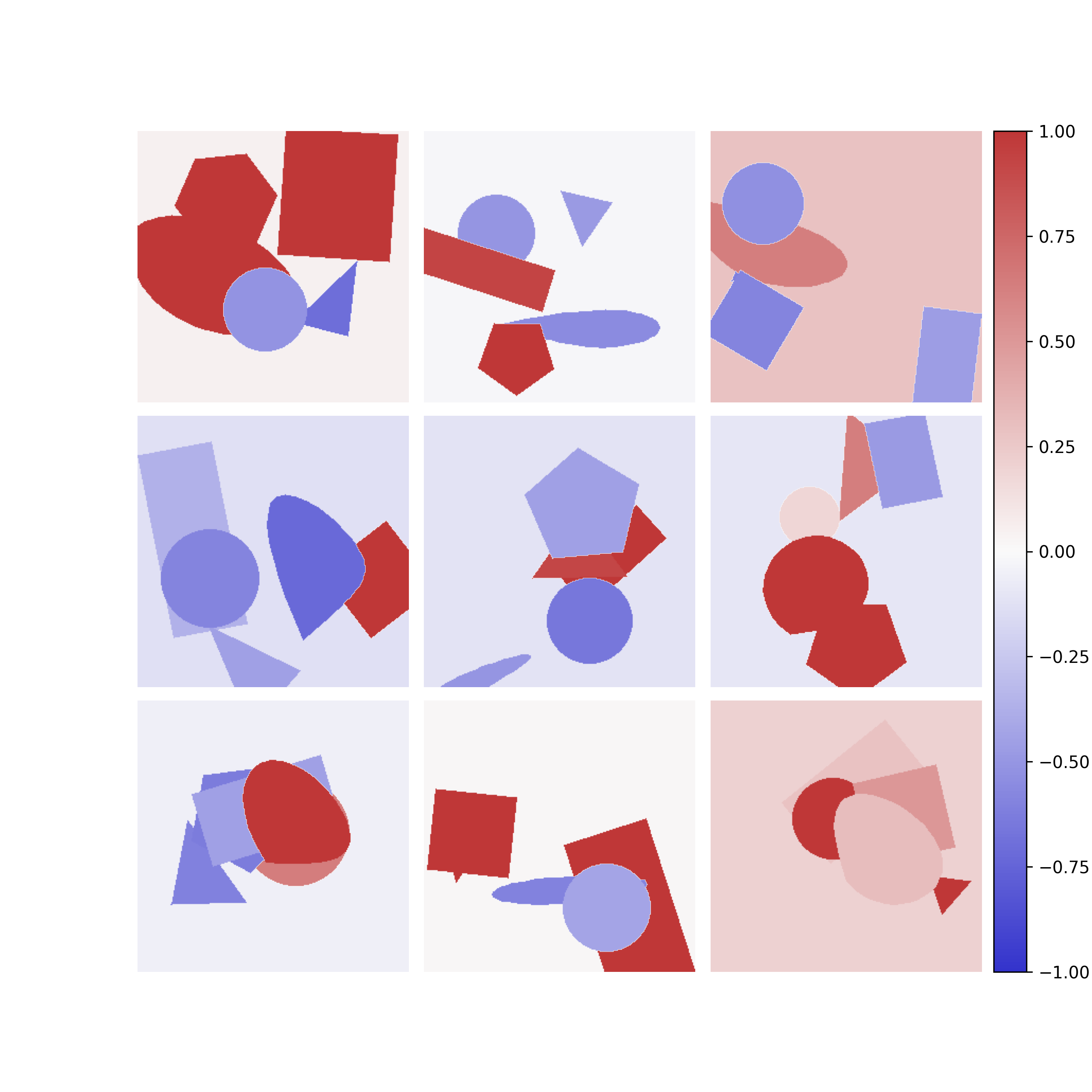}
    \caption{Samples in 2D shape datasets for pretraining the diffusion model.}
    \label{fig.ddpm_data}
\end{figure}

A total of 50,000 samples of conductivity distributions were generated, normalized to the range of $[-1, 1]$, and used as input for training the diffusion model. The model was trained using the Improved-DDPM \cite{improved_ddpm} architecture and its hybrid loss function. Training was conducted on an NVIDIA GeForce RTX 4090 GPU for 50,000 epochs with a batch size of 128. The training process is detailed in Algorithm \ref{alg.ddpm_data}, and sample dataset images are shown in Figure \ref{fig.ddpm_data}.

\begin{algorithm}[H]
\caption{Generating 2D Shape Datasets for pretraining the diffusion model}
\label{alg.ddpm_data}
\begin{algorithmic}[1]
\STATE \textbf{Input:}
\begin{itemize}
    \item Total number of figures: $N_{total}$
    \item Number of objects per figure: $N_{objects}$
    \item Location limits: $x_{min}$, $x_{max}$, $y_{min}$, $y_{max}$
    \item Size limits: $l_{min}$, $l_{max}$
    \item Scale factor: $A$
    \item Means and standard deviations for conductivity: $\mu_{high}$, $\mu_{low}$, $\mu_{bkg}$, $\sigma_{high}$, $\sigma_{low}$, $\sigma_{bkg}$
\end{itemize}

\FOR{$i = 1$ \TO $N_{total}$}
    \STATE Sample background conductivity: $\sigma_{bkg} \sim \left|\mathcal{N}(\mu_{bkg}, \sigma_{bkg})\right|$
    \STATE Initialize new image with dimensions $256 \times 256$ and fill with background value $\sigma_{bkg}$
    \FOR{$j = 1$ \TO $N_{objects}$}
        \STATE Randomly select shape type: \{Bezier curve, triangle, circle, polygon, rectangle\}
        \STATE Randomly select conductivity type: \{high, low\}
        \STATE Sample random location: $(x, y) \sim \mathcal{U}(x_{min}, x_{max}), \mathcal{U}(y_{min}, y_{max})$
        \STATE Sample random size: $l \sim \mathcal{U}(l_{min}, l_{max})$
        \IF{conductivity == low}
            \STATE Sample object conductivity: $\sigma_{object} \sim \left|\mathcal{N}(\mu_{low}, \sigma_{low})\right|$
        \ELSE
            \STATE Sample object conductivity: $\sigma_{object} \sim \left|\mathcal{N}(\mu_{high}, \sigma_{high})\right|$
        \ENDIF
        \STATE Draw the object with the selected shape, location, size, and conductivity
    \ENDFOR
    \STATE Normalize image: $\text{img} \gets \left(\frac{\text{img}}{A}\right) \times 2 - 1$
\ENDFOR
\end{algorithmic}
\end{algorithm}

\subsection{Bandwidth Control Mechanism}
\label{sec.inr_for_eit}
In the INR-based EIT reconstruction framework, maximizing the utility of measured voltage information hinges on the INR's ability to accurately reflect the data, necessitating a high bandwidth \cite{rethink_pe}. However, FE meshes usually exhibit relatively low information density, which demands that the INR possess robust generalization abilities. Consequently, the bandwidth cannot be set too high.
Moreover, the information density of FE meshes varies with different degrees of discretization, meaning that the optimal bandwidth must adapt when the mesh changes. To improve the INR-based method's robustness to varying mesh discretizations, we propose a bandwidth control mechanism. 
This mechanism defines the bandwidth parameter $b$ as follows:
\begin{equation} 
\label{eq.bandwidth} b \propto \frac{1}{d} = k \sqrt{N_e}, 
\end{equation}
where $d$ is the average edge length of the mesh elements, $N_e$ represents the total number of  elements, and the hyperparameter $k$ adjusts the bandwidth scaling according to the mesh properties. This approach enables the INR to dynamically adjust its representation capabilities in response to changes in mesh discretization.

\subsection{Execution and Optimization of the Proposed Approach}
The diffusion model, which is pre-trained on 2D shape images, remains fixed throughout the reconstruction process, as indicated by the lock symbol in the architecture of the proposed approach (see Fig.~\ref{fig.structure}).
During this process, the only component that is updated iteratively is the MLP $F_\theta$, which serves as the implicit representation. This MLP is composed of four fully connected layers, each with 128 neurons. Input coordinates are encoded using RFF with a sampling count of $n = 128$. The sampling bandwidth $b$ is determined by equation (\ref{eq.bandwidth}) with $k = 1.5\times 10^{-2}$. The optimization process involves using the Jacobian matrix of the forward problem along with neural network gradients obtained via backpropagation. MLP parameters $\theta$ are updated using the Adam optimizer with a learning rate of 0.01.

Due to the high CPU load associated with computing the forward problem and Jacobian matrix, the reconstruction process is optimized for performance by executing the diffusion model and computing Jacobian on an NVIDIA GeForce RTX 4070 GPU. In contrast, all other computations are executed on an Intel Core i7-12700K CPU.

\subsection{Evaluation Metrics}
To evaluate the quality of results from simulated data, we use two metrics: the Structural Similarity Index (SSIM) \cite{ssim_2004tip} and the Peak Signal-to-Noise Ratio (PSNR). PSNR is expressed in decibels (dB), while SSIM is a ratio ranging from 0 to 1. Higher values for both metrics indicate better alignment with the original image.

For experimental data reconstruction, the accuracy of SSIM and PSNR measurements cannot be assured due to the lack of precise registration; only a photograph of the experimental tank is available. Given that the experiment involves only a few discrete, regular geometric objects, visual assessment is used to gauge the accuracy of shape reconstruction. To evaluate properties of the recovered inclusions, we calculate the Relative Size Coverage Ratio (RCR) \cite{dong2020tmi}:

\begin{equation}
    \label{eq.RCR}
    \text{RCR} = \frac{\text{CR}}{\text{CR}_{\text{true}}},
\end{equation}
where CR represents the ratio of the reconstructed area of inclusions to the measurement domain, and $\text{CR}_{\text{true}}$ is the CR of the actual inclusion.
A value of 1 indicates an exact match between the true and estimated areas of inclusion, while values greater or less than 1 signify overestimation or underestimation, respectively.

\begin{table*}[]
\centering
\setlength{\tabcolsep}{7pt}
\renewcommand{\arraystretch}{1.2}
\caption{Evaluation criterion RCR of the experimental studies}
\label{tab.results}
\begin{threeparttable}
  \begin{tabular}{c|cc|cc|ccc|ccc}
    \cline{1-11}
    \multicolumn{1}{c|}{} &\multicolumn{2}{c|}{Case 1} & \multicolumn{2}{c|}{Case 2} & \multicolumn{3}{c|}{Case 3} & \multicolumn{3}{c}{Case 4}\\
    \multicolumn{1}{c|}{}& RCR$_\text{Lt}$ & RCR$_\text{r}$ & RCR$_\text{r}$ & RCR$_\text{c}$ & RCR$_\text{Lt}$ & RCR$_\text{r}$ & RCR$_\text{c}$ & RCR$_\text{St}$ & RCR$_\text{r}$ & RCR$_\text{c}$ \\
    \cline{1-11}
    TV      & 1.08 & 0.92 & 1.02 & 0.86 & 1.13 & 1.28 & 0.75 & 1.08 & 0.99 & 0.84 \\
    INR+TV  & 1.19 & 1.15 & 1.13 & 1.09 & 1.20 & 1.10 & 1.02 & 0.96 & 1.11 & 1.08 \\
    Diff-INR& 1.12 & 1.09 & 1.07 & 1.01 & 1.07 & 1.05 & 1.00 & 1.06 & 1.06 & 1.06 \\
    \cline{1-11}
   \end{tabular}
\begin{tablenotes} 
\item The subscript `Lt' indicates the large triangle-shaped inclusion in Cases 1 and 3, `St' represents the small triangle-shaped inclusion in Case 4, `r' stands for the rectangle, and `c' corresponds to the circle.

\end{tablenotes}
\end{threeparttable}
\end{table*}

\begin{figure*}[!t]
\centering
\includegraphics[trim={0 80 10 55},clip,width=6.4in]{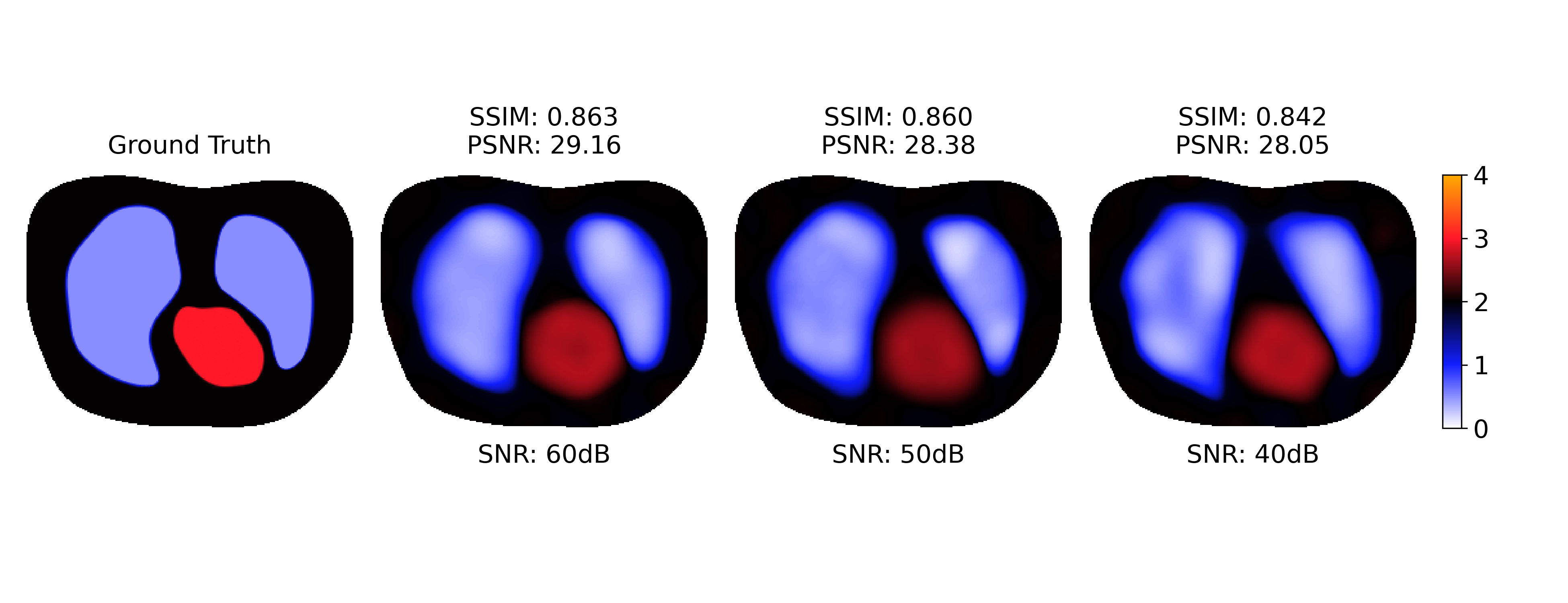}
\caption{Simulation results using a human thorax phantom under different noise levels.}
\label{fig.noise_level}
\end{figure*}

\section{Results and Discussion}
\label{sec_eval}
In this section, we thoroughly evaluate the proposed Diff-INR method compared to other representative reconstruction techniques. This analysis is designed to assess the effectiveness of the generative regularization approach and the overall performance of the Diff-INR framework. 
Given that the proposed approach is a self-supervised method, we do not include comparisons with supervised approaches. Instead, we compare it with classical Total Variation (TV) based reconstruction \cite{TV_pdipm_2010tmi} and recent unsupervised neural network-based reconstruction algorithms, specifically INR \cite{junwu2023inr_eit}. For clarity, we refer to INR+TV as INR-based reconstruction with TV regularization.

\subsection{Simulation results and Robustness to Noise Levels}
To assess the performance of the Diff-INR method under realistic conditions, we conducted a series of numerical simulations using a human thorax phantom, generating measurement data across a range of signal-to-noise ratios (SNR). This setup aims to emulate the complexities encountered in clinical EIT applications.

As depicted in Figure \ref{fig.noise_level}, the Diff-INR method demonstrates significant robustness across various noise levels, with SNR values ranging from 60 to 40 dB. The reconstructed lung contours maintain a high degree of accuracy despite variations in noise conditions. As noise levels increase, the impact of the regularization term becomes more pronounced. Specifically, in the 40 dB SNR scenario, the reconstructed lung contours adopt a polygonal shape with relatively straight edges, a feature that helps to counteract the detrimental effects of elevated noise. It is important to note that EIT's inherent lower sensitivity in central regions, as opposed to near-boundary areas, combined with the shielding effect of the lungs over the heart, makes it particularly challenging to accurately reconstruct the heart's shape. Nevertheless, the generative regularization in Diff-INR effectively constrains the heart's representation to a polygonal shape approximating a circle, ensuring a relatively precise positional estimate.

Moreover, this human thorax scenario also serves as an out-of-distribution (OOD) test relative to the simple 2D shape dataset used during training. The high reconstruction quality achieved in this context underscores the strong generalization capabilities of the generative regularization embedded in Diff-INR. Unlike many conventional models, which often struggle when confronted with OOD samples, Diff-INR successfully extrapolates beyond the training distribution, delivering accurate reconstructions even when applied to target shapes and conditions not encountered during the training phase. This ability to generalize effectively to OOD samples highlights Diff-INR's potential for robust application in diverse clinical settings, where the imaging target can vary significantly from the training data.

\begin{figure*}[!t]
\centering
\includegraphics[trim={0 0 0 0},clip,width=6in]{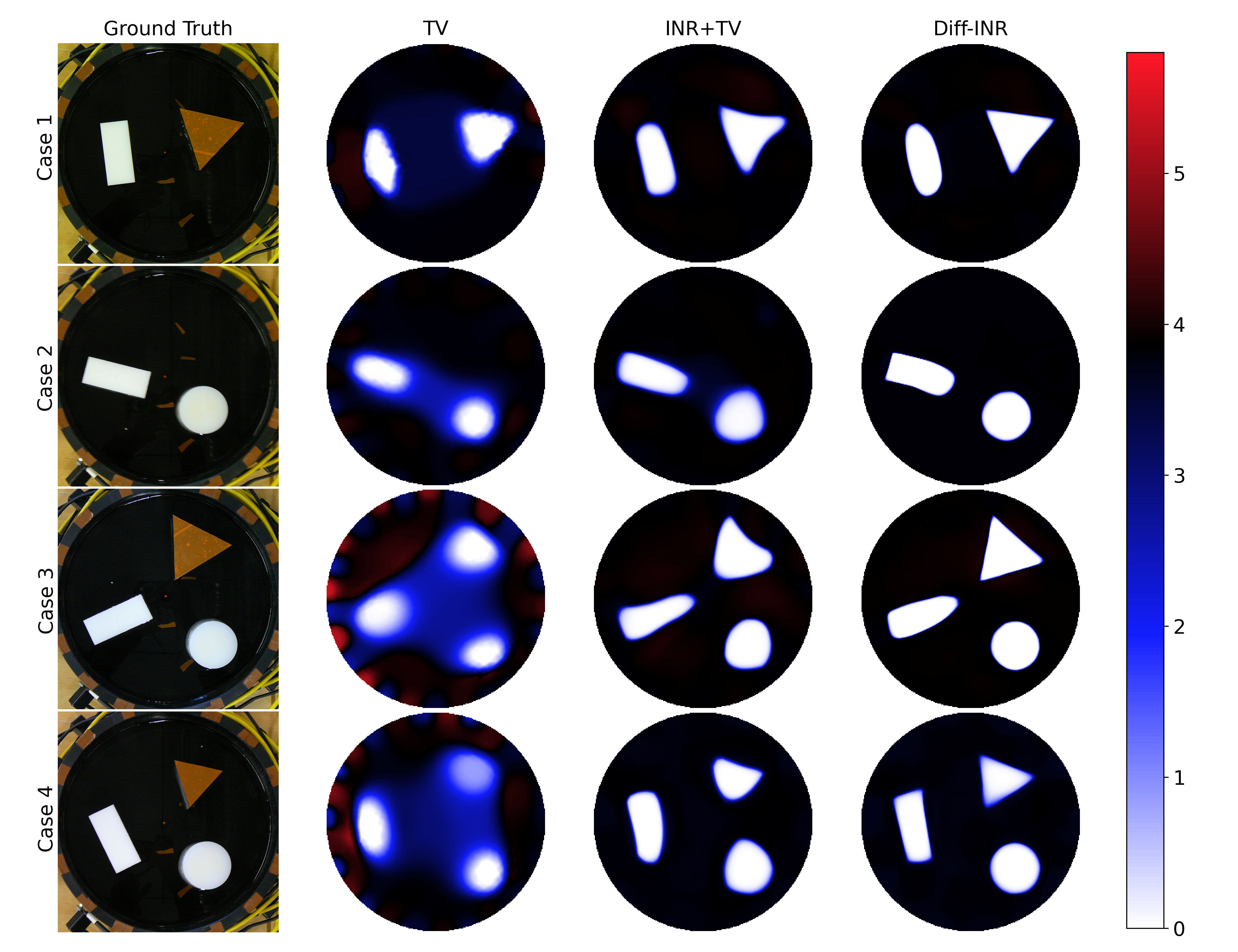}
\caption{Results of experimental data reconstruction using TV, INR+TV and Diff-INR.}
\label{fig.compare_methods}
\end{figure*}

\subsection{Experimental results}
Figure \ref{fig.compare_methods} illustrates the reconstructions obtained using water tank data. The staircase effect observed in TV-based reconstructions is anticipated, as TV regularization inherently promotes sharp transitions to minimize total variation. This results in distinct blocky artifacts. In contrast, INR+TV demonstrates significant improvements over TV, offering clearer boundaries, better artifact suppression, cleaner backgrounds, and more accurate object shapes. Nonetheless, INR+TV still exhibits some deformation in specific instances, such as triangular-shaped inclusions (refer to Figure \ref{fig.compare_methods}).

The proposed Diff-INR method effectively addresses the limitations encountered with the aforementioned approaches. It produces well-defined boundaries, accurate shapes, positions, and sizes, along with an exceptionally clean background. As shown in Figure \ref{fig.compare_methods}, Diff-INR excels in reconstructing fine-grained object structures, including sharp corners of polygonal shapes and perfect circles. This superior performance is further corroborated by Table \ref{tab.results}, where the RCR for Diff-INR approaches the ideal value of 1, reflecting the highest reconstruction accuracy, particularly for circular objects where the results are notably precise. These observations indicate that Diff-INR not only surpasses other methods in visual quality but also delivers the most accurate quantitative reconstruction outcomes.

The effectiveness of Diff-INR is largely attributed to its application of multi-scale regularization through varying time steps $t$ in the diffusion model. Larger values of $t$ correspond to higher noise levels and less retention of original image details, which helps to mask low-contrast artifacts within the noise. This approach allows for more uniform conductivity across large areas, resulting in a cleaner background free of staircase-like artifacts. The impact of diffusion time steps $t$ on reconstruction accuracy will be further examined in the following subsection.

In summary, the diffusion model-based regularization term in Diff-INR not only maintains the advantages of INR reconstruction but also provides two additional critical benefits: it significantly reduces the appearance of artifacts and ensures a clean background while offering effective geometric shape guidance for accurate reconstruction.

\begin{figure*}[!t]
\centering
\includegraphics[trim={0 0 0 0},clip,width=6.2in]{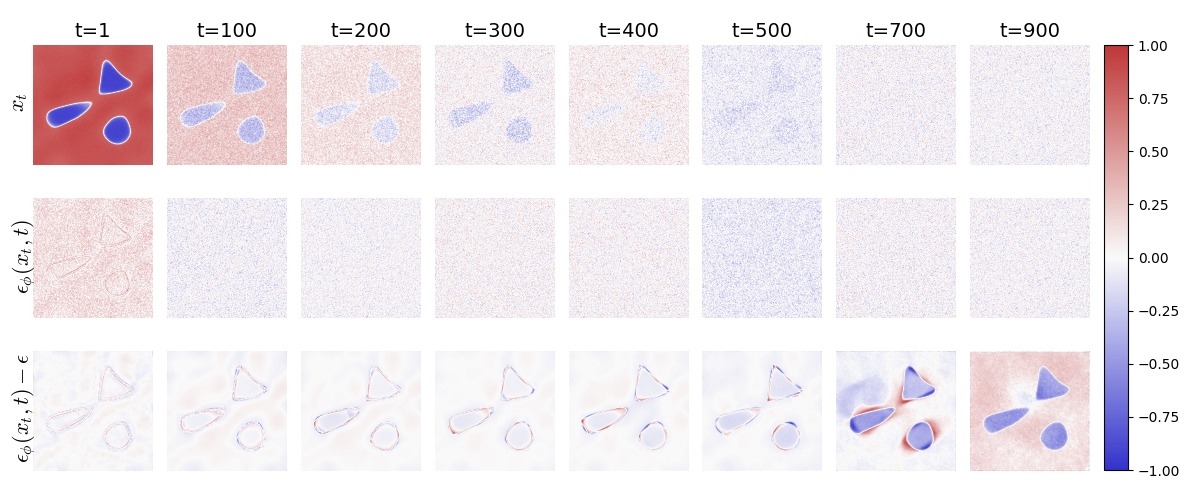}
\caption{Image corresponding to $x_t$, $\epsilon_\phi(x_t, t)$, and $[\epsilon_\phi(x_t, t) - \epsilon]$ across various diffusion time steps.}
\label{fig.diff_reg_timestep}
\end{figure*}

\subsection{Exploring Diffusion Time Step Variations and Their Influence on EIT Reconstruction }
To maintain coherence with the prior discussion, we begin by considering the robustness of the proposed approach concerning the diffusion time steps $t$. According to Equation \ref{eq.L_reg_0}, the generative regularization theoretically minimizes $\mathcal{L}_\text{reg}$ across all time steps. However, we found that, unlike other studies that sample across the full range from $T$ to 0 steps, better results can be achieved by selecting a subset of time steps for stochastic optimization.

To investigate this, we analyzed the impact of generative regularization at different time steps on the reconstruction process of Case 3 (see Figure \ref{fig.diff_reg_timestep}). 
The first row depicts the noisy reconstructed images $x_t$ at various time steps. 
As expected, the original image's semantic information diminishes with increasing $t$. 
The second row of Figure \ref{fig.diff_reg_timestep} illustrates the output of the diffusion model $\epsilon_\phi(x_t, t)$, while the third row shows the effect of diffusion regularization, obtained by subtracting the random Gaussian noise $\epsilon$ from $\epsilon_\phi(x_t, t)$. 
The magnitude of $\epsilon_\phi(x_t, t)-\epsilon$ influences the gradient updates during optimization. 
Grayer regions, where the color bar value is close to zero, represent smaller update magnitudes, whereas more vibrant regions indicate larger update magnitudes. This highlights the areas with the most significant conductivity changes during the optimization step. 

When $t$ is small, the sensitive regions are relatively small and detailed, corresponding to high spatial frequencies. As $t$ increases, the sensitive regions expand, representing lower spatial frequencies. This observation suggests that generative regularization dynamically adapts its influence on the reconstruction process at different time steps.
Interestingly, at intermediate time steps (e.g., $t=400$ in Figure \ref{fig.diff_reg_timestep}), the sensitive regions concentrate mainly on the edges of the object, while the background and interior remain relatively insensitive. This behavior effectively aids in EIT reconstruction, where accurately obtaining structural edges is critical. By focusing regularization on edge regions, the method facilitates a more precise reconstruction of object contours, leading to accurate interior values.

However, at excessively large time steps, such as $t=900$, the lack of informative content in $x_t$ causes the main object in the reconstructed image to become a sensitive region, suggesting that the regularization might distort the current structure rather than refine it. This underscores the importance of carefully selecting the time step range to balance structural preservation and regularization effects.

Unlike diffusion sampling methods \cite{DPS}, \cite{reddiff}, \cite{diff_fastMRI_2023tmi}, the Diff-INR approach does not perform a complete diffusion sampling.
Given the inherent instability of EIT reconstruction, determining the reverse diffusion time schedule based solely on iteration steps is impractical. Instead, each $x_t$ at different time steps $t$ is recalculated from the current reconstruction. This strategy enables the generative regularization to apply guidance corresponding to various spatial frequencies at different time steps, ensuring effective support within the frequency range of interest for the reconstruction.
Additionally, due to the ill-posed nature of EIT, the iteration process is prone to converging to local optima. To mitigate this, we introduce a certain degree of perturbation to the reconstructed structure at each iteration step by randomly selecting the time $t$. This approach assists the optimization process in escaping local optima and facilitates a more comprehensive exploration of the solution space.
Therefore, we adopt a stochastic optimization approach, randomly sampling $t$ from a uniform distribution $\mathcal{U}[t_{min},t_{max}]$. Based on trial and error, we found that $t \sim \mathcal{U}[100,400]$ yields good results in most cases.

Interestingly, we also observed that fixing $t$ to a specific value, such as $t=300$, can yield more accurate results in simpler cases. This likely occurs because the structural complexity associated with $t=300$ may better match the characteristics of these particular test cases.
However, this setting is not universally effective, and determining the optimal fixed $t$ without prior knowledge of the ground truth remains challenging. These findings suggest that future research could further explore time step selection, potentially developing adaptive strategies to automatically determine the most suitable $t$ for specific problem settings. Such strategies could help maximize the performance of generative regularization. Nonetheless, to maintain a broadly applicable framework, we have retained the random optimization approach in this work, as it ensures robust performance across a diverse range of problem instances without requiring extensive problem-specific tuning.

\begin{figure*}[!t]
  \centering
  \includegraphics[trim={0 0 0 0},clip,width=6.4in]{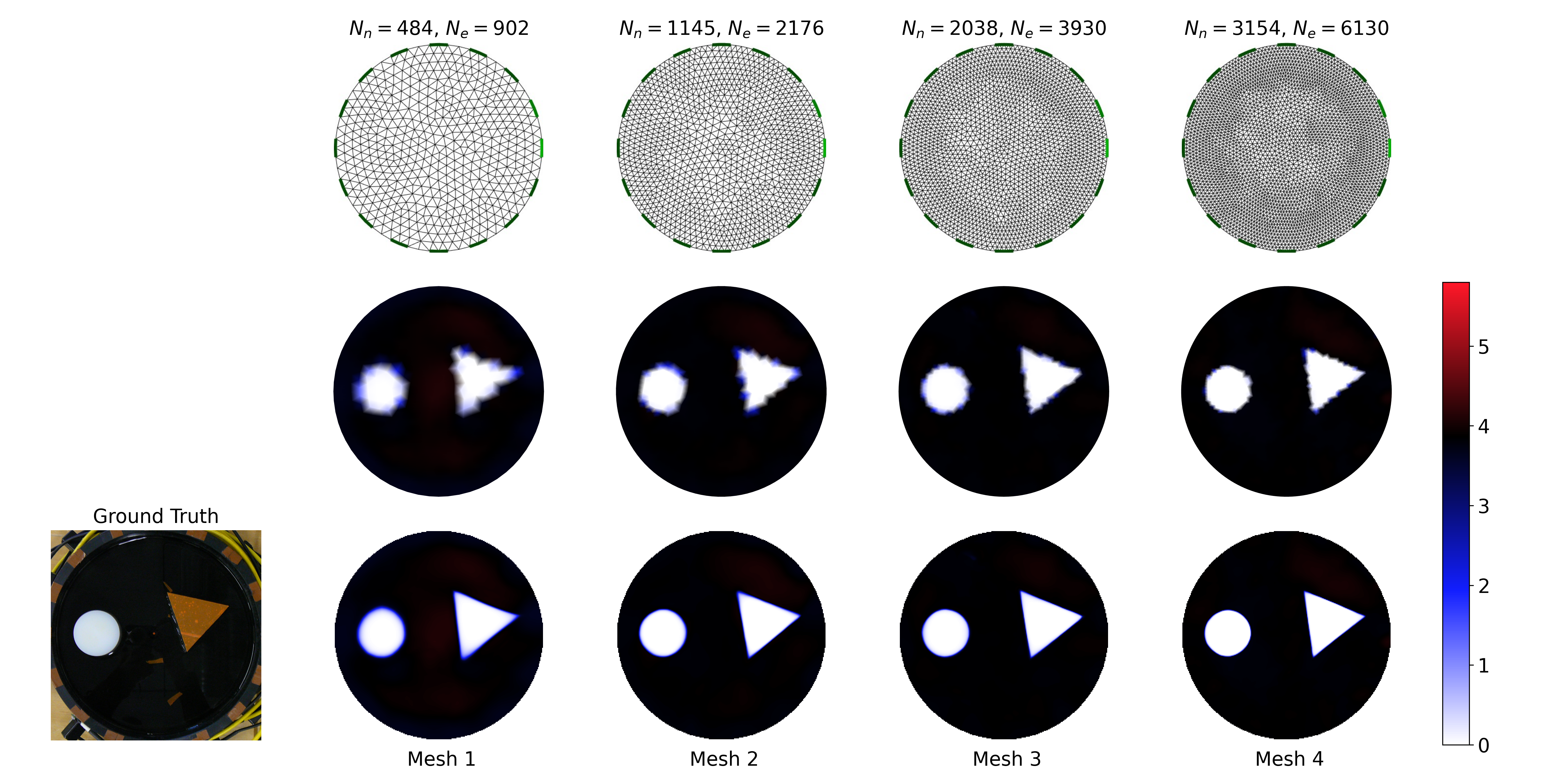}
  \caption{Diff-INR-based reconstruction with various inverse meshes. The first row shows the  meshes with different discretization levels. The second row shows the conductivity distribution on these meshes, while the third row presents the conductivity distribution in the pixel domain.}
  \label{fig.mesh_level}
  \end{figure*}
  
\subsection{Robustness of Diff-INR to Finite Element Mesh Variations and Hyperparameter Selection}
To achieve reliable reconstruction results in EIT, it is crucial to carefully consider the discretization level, mesh quality, and consistency between the forward and inverse models. Given the sensitivity of EIT reconstruction to FE mesh discretization levels, selecting an appropriate discretization is vital for balancing both accuracy and computational efficiency.

To assess the robustness of the proposed Diff-INR approach, we evaluated its performance across a variety of finite element meshes with different discretization levels, labeled as Mesh 1 to Mesh 4, where higher numbers indicate finer mesh density.
As illustrated in Figure \ref{fig.mesh_level}, Diff-INR consistently delivers high-quality reconstruction results across these varying discretization levels. This robustness can be attributed to the bandwidth control mechanism described in Section \ref{sec.inr_for_eit}.
Notably, the reconstructions for each FE mesh are nearly identical, indicating that Diff-INR can be effectively applied even to relatively sparse meshes. This adaptability is particularly advantageous for practical applications, as it allows for significant computational savings when using coarser meshes.
Additionally, the INR-based method achieves accurate and continuous conductivity representation without requiring interpolation, thus enabling super-resolution capabilities beyond the native resolution of the FE mesh. For instance, the third row of Figure \ref{fig.mesh_level} is derived from the pixel domain, while the second row of Figure \ref{fig.mesh_level} is the conductivity distribution plotted on FE mesh.

\begin{figure}
    \centering
    \includegraphics[trim={2 0 10 10},clip,width=3.5in]{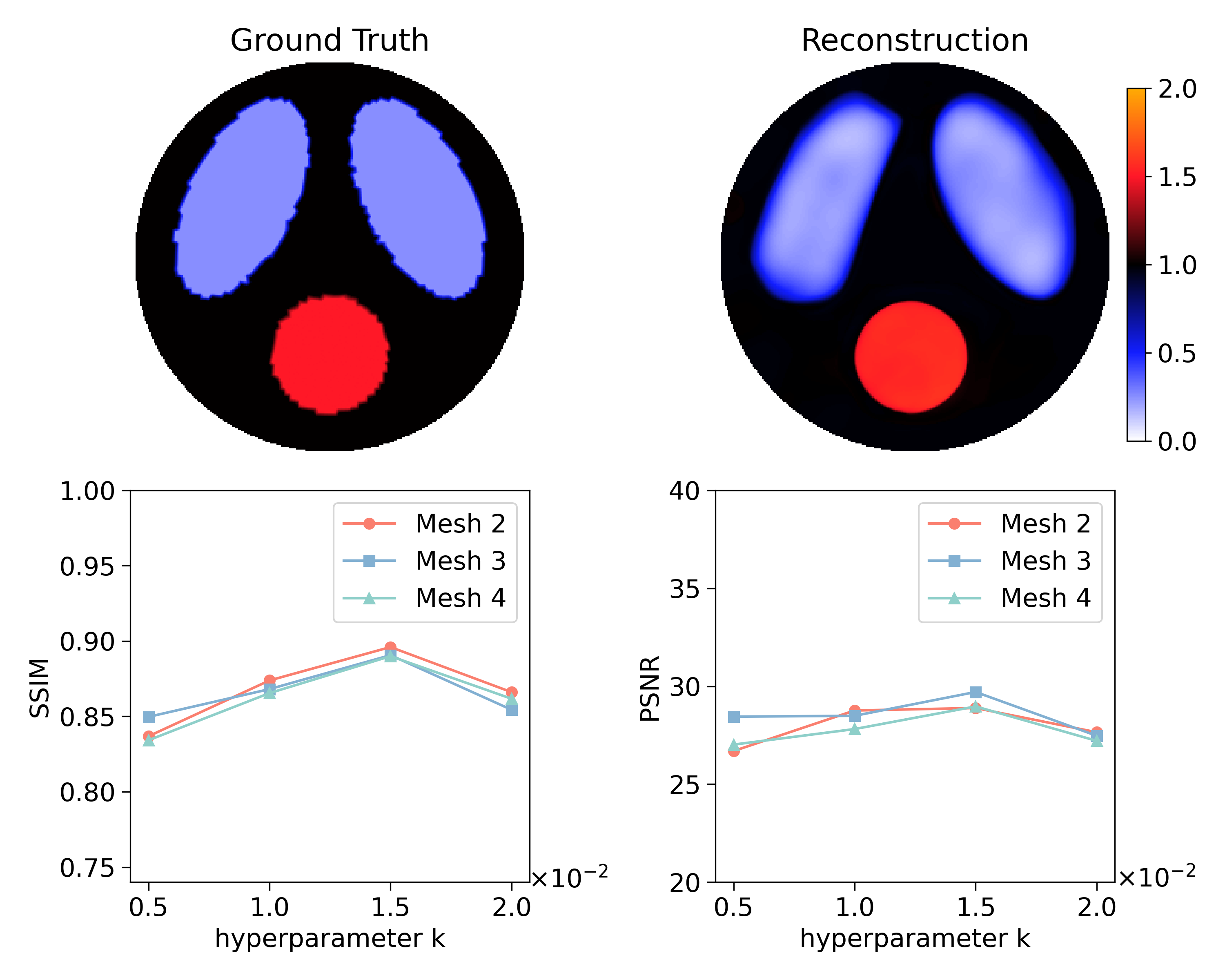}
    \caption{PSNR and SSIM metrics for varying hyperparameter $k$ in determining the bandwidth $b$. The first row displays the simulation case and a representative reconstruction.}
    \label{fig.ssim_psnr}
\end{figure}

To further investigate the robustness of the proposed approach concerning the hyperparameter $k$ in determining the bandwidth $b$, as given in Equation \ref{eq.bandwidth}, we performed a series of reconstructions in a simulation study using three different inverse meshes, as labeled Mesh 2 to Mesh 4 in Figure \ref{fig.mesh_level}.
For simplicity, we present only the evaluation metrics and one representative reconstructed image for $k = 1.5 \times 10^{-2}$ using Mesh 3, as illustrated in Figure \ref{fig.ssim_psnr}.

Our observations indicate that for a given value of $k$, the reconstruction results across various meshes are consistently similar. This suggests that $k$ accurately reflects the mesh discretization level and effectively mitigates the impact of varying mesh resolutions on reconstruction quality.
Moreover, within a specific range, increasing the bandwidth can lead to improved results, as demonstrated by $k = 1.5 \times 10^{-2}$, which yields the highest SSIM values. However, selecting an excessively large bandwidth can hinder the INR's ability to generalize and accurately represent positions not aligned with the FE nodes, ultimately leading to reduced performance.

In summary, Diff-INR shows strong performance across various finite element mesh discretizations and hyperparameter choices, demonstrating its robustness and versatility in EIT reconstruction tasks. This flexibility not only ensures high-quality reconstructions but also provides practical benefits like computational efficiency and enhanced resolution capabilities.

\section{Conclusion}
\label{sec_conclusion}
This paper introduced the Diff-INR method for EIT reconstruction, which combines diffusion-based regularization with Implicit Neural Representation to significantly enhance accuracy and reduce artifacts. The method demonstrates high-quality results even with sparse mesh discretizations, offering practical benefits by improving computational efficiency and enabling effective use of coarser meshes.
Our analysis confirms that Diff-INR maintains robust performance across various mesh densities, which is critical for real-world applications where computational resources may be limited. The method's adaptability to different hyperparameter settings, particularly the bandwidth coefficient $k$ and diffusion time step $t$, further enhances its effectiveness. The use of stochastic optimization for selecting $t$ allows dynamic adjustment to spatial frequencies, which improves reconstruction quality.

In summary, Diff-INR represents a notable advancement in EIT reconstruction, delivering robust, high-resolution results with practical advantages for diverse scenarios. Future work should focus on optimizing hyperparameter selection and exploring clinic applications to further enhance EIT reconstruction techniques.
While this study concentrated on 2D EIT reconstruction, Diff-INR can be extended to 3D EIT, where partial sections of the 3D domain can be used to compute regularization loss and guide optimization. Moreover, the generative regularization approach could benefit from incorporating more advanced foundation models, such as Stable Diffusion, to integrate text-guided inputs for improved reconstruction.

\section*{Acknowledgment}
The authors would like to thank Mr. Hongyan Xiang from the University of Science and Technology of China for his valuable discussions on deriving the idea of applying diffusion models to inverse problems.

\bibliographystyle{unsrt}  
\bibliography{references}  

\end{document}